\documentclass[twocolumn,secnumarabic,groupaddress,nofootinbib,aps,pra,showpacs,showkeys]{revtex4-1}

\usepackage{amsthm}
\usepackage{amsmath}
\usepackage{amssymb}
\usepackage{color}
\usepackage{adjustbox}
\usepackage{graphicx}
\usepackage{hyperref}
\usepackage{multirow}
\usepackage{array}
\usepackage{blindtext}
\usepackage{hyperref}
\usepackage{epstopdf}
\usepackage{array,multirow}
\usepackage{threeparttable}
\usepackage{latexsym}
\usepackage{amssymb,amsmath}
\usepackage{color}
\usepackage{url}
\usepackage{ulem}

\begin{document}
\title{Entanglement distribution over 150\,km in wavelength division multiplexed channels for quantum cryptography}
\author{Djeylan Aktas}
\affiliation{Universit\'e Nice Sophia Antipolis, Laboratoire de Physique de la Mati\`ere Condens\'ee, CNRS UMR 7336, Parc Valrose, 06108 Nice Cedex 2, France}
\author{Bruno Fedrici}
\affiliation{Universit\'e Nice Sophia Antipolis, Laboratoire de Physique de la Mati\`ere Condens\'ee, CNRS UMR 7336, Parc Valrose, 06108 Nice Cedex 2, France}
\author{Florian Kaiser}
\affiliation{Universit\'e Nice Sophia Antipolis, Laboratoire de Physique de la Mati\`ere Condens\'ee, CNRS UMR 7336, Parc Valrose, 06108 Nice Cedex 2, France}
\author{Tommaso Lunghi}
\affiliation{Universit\'e Nice Sophia Antipolis, Laboratoire de Physique de la Mati\`ere Condens\'ee, CNRS UMR 7336, Parc Valrose, 06108 Nice Cedex 2, France}
\author{Laurent Labont\'e}
\affiliation{Universit\'e Nice Sophia Antipolis, Laboratoire de Physique de la Mati\`ere Condens\'ee, CNRS UMR 7336, Parc Valrose, 06108 Nice Cedex 2, France}
\author{S\'ebastien Tanzilli}\email{sebastien.tanzilli@unice.fr}
\affiliation{Universit\'e Nice Sophia Antipolis, Laboratoire de Physique de la Mati\`ere Condens\'ee, CNRS UMR 7336, Parc Valrose, 06108 Nice Cedex 2, France}

\date{\today}



\begin{abstract}
Granting information privacy is of crucial importance in our society, notably in fiber communication networks. Quantum cryptography provides a unique means to establish, at remote locations, identical strings of genuine random bits, with a level of secrecy unattainable using classical resources.
However, several constraints, such as non-optimized photon number statistics and resources, detectors' noise, and optical losses, currently limit the performances in terms of both achievable secret key rates and distances.
Here, these issues are addressed using an approach that combines both fundamental and off-the-shelves technological resources. High-quality bipartite photonic entanglement is distributed over a 150\,km fiber link, exploiting a wavelength demultiplexing strategy implemented at the end-user locations. It is shown how coincidence rates scale linearly with the number of employed telecommunication channels, with values outperforming previous realizations by almost one order of magnitude. Thanks to its potential of scalability and compliance with device-independent strategies, this system is ready for real quantum applications, notably entanglement-based quantum cryptography.\\
\begin{center}
\includegraphics[width=92mm]{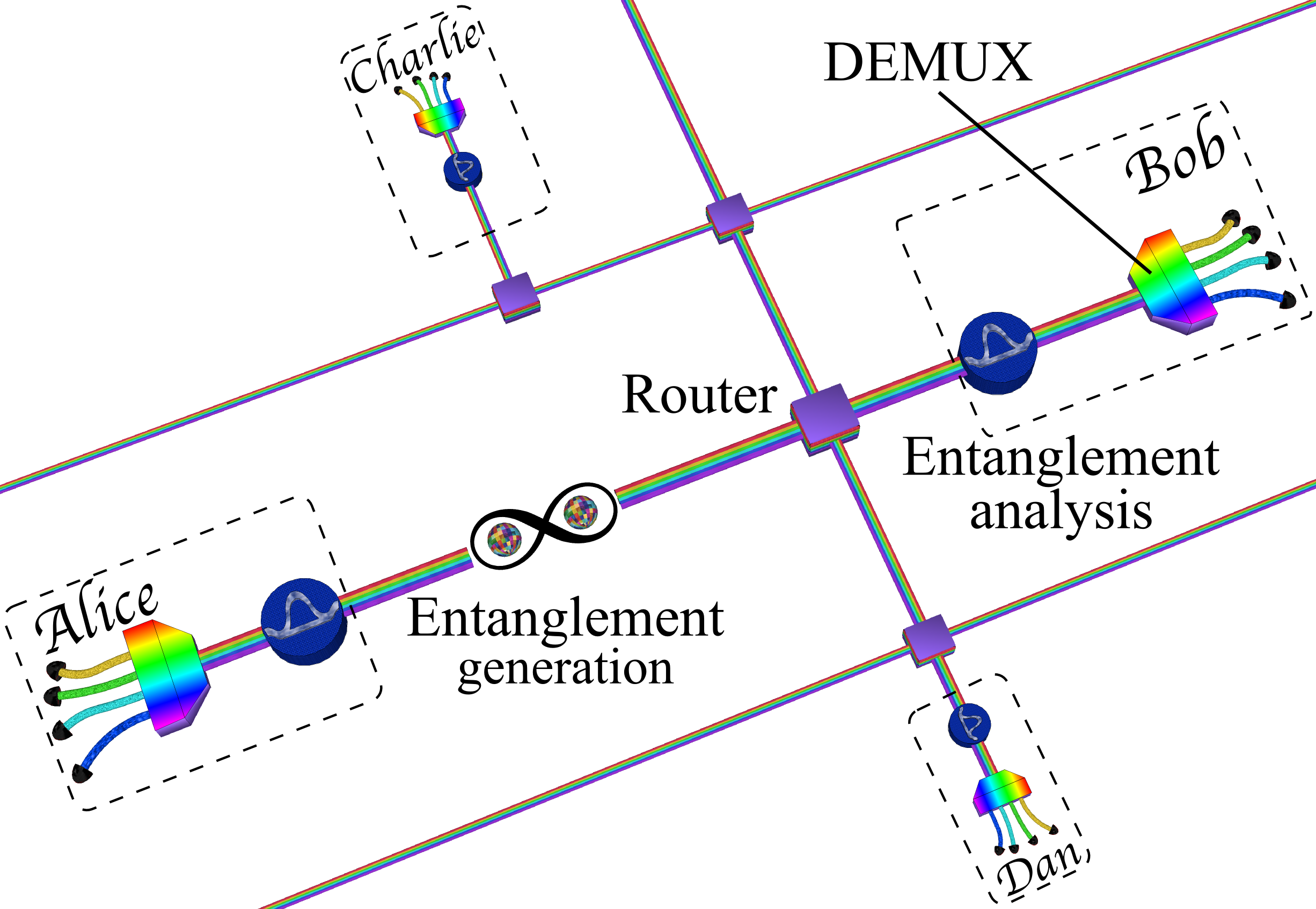}
\end{center}
\end{abstract}

\maketitle

\section{Introduction}

Entanglement lies at the heart of quantum communication and computing protocols~\cite{Brunner_Bell_14} as it offers the unique possibility of building device independent quantum systems~\cite{Vazirani_PRL_14,Aktas_LDL_2016,Hensen_LoopholeFree_15}.
Similarly to today's high-performance classical optical networks, future quantum networks will be based on several entanglement-based key building blocks in which photonics technologies are expected to play a major role~\cite{Tanzilli12_LPRreview}. On one hand, reliable terminal stations, connected by optical fiber links, can serve either to generate and launch quantum information into distribution channels, or to receive and analyze it. On the other hand, quantum repeaters~\cite{Simon07} and amplifiers~\cite{AnthoAmpli13} are used to overcome losses in these channels towards increasing the overall network efficiency.
As in current telecommunication systems, quantum bit rates can be further increased by multimodal operation, \textit{i.e.}, the ability to spatially~\cite{Meany_LPR14}, temporally~\cite{Ngah_LPR15}, and/or spectrally~\cite{Lim10} multiplex many independent signals into a single communication channel and to appropriately demultiplex them at various locations. This allows both exploiting the maximum capacity of a given channel, and routing information to different users dynamically.

So far, quantum communication systems have addressed several, however not all, of these building blocks. There exist a variety of high-performance terminal stations, such as those intended to code quantum information onto single or entangled photons.
Notably, entangled photon-pair distribution has already been achieved over 300\,km fiber links~\cite{Takesue13_300km}.
Furthermore, quantum relay and repeater stations have also been demonstrated in the laboratory, relying on two-photon interference~\cite{Kaiser_IEEE14}, and on light-matter interfaces with cold atomic ensembles~\cite{SimonRepeater11} or ion-doped crystals~\cite{Riedmatten10}.
Up to now, signal multiplexing/demultiplexing and information routing have not been fully exploited for quantum network applications, such that fiber distribution channels are currently operated far below their capacity~\cite{Meany_LPR14,Ngah_LPR15,Lim10}.

In this context, we report in the following an advanced entanglement distribution scheme~\cite{Ekert91}, that successfully merges a high-quality two-photon generator~\cite{TheKaiser14Long} and high-performance dense wavelength division multiplexing (DWDM) components coming from telecommunication technologies.
Such a strategy enables increasing photon pair coincidence rates by almost one order of magnitude at a distance of 150\,km, and can be straightforwardly applied to quantum cryptography (QC) tasks.

\section{Experimental section \& methods}

\subsection{A two-user demultiplexing scenario for high bit rate QC}
Consider a single source that generates a flux of wavelength correlated photon pairs, as can be obtained with energy-time entangled photons emitted via the process of spontaneous parametric down-conversion (SPDC) in a nonlinear optical crystal. Here, conservation of the energy implies that pairs of correlated photons are always produced symmetrically with respect to the center of the emission spectrum.
As shown in Fig.~\ref{fig_spectrum}, if the spectrum is broad enough, e.g., covering the entire telecom C-band, two main wavelength-demultiplexing scenarios can be implemented experimentally. First, the spectrum can be demultiplexed for entangling multiple pairs of users when supplied with pairs of photons having complementary wavelengths. Alternatively, or additionally, each pair of users can further exploit local demultiplexing to obtain a bit rate increase equal to the number of demultiplexed pairs of wavelength-correlated channels.
In the following, we demonstrate the realization of a two-user scenario, as shown in Fig.~\ref{fig_setup}, where both users demultiplex their signals using an 8$\times$100\,GHz bandwidth DWDM, belonging to the grid of the international telecommunication union (ITU).

\subsection{Generating broadband energy-time entangled photons}

\begin{figure}
\centering
\includegraphics[width=75mm]{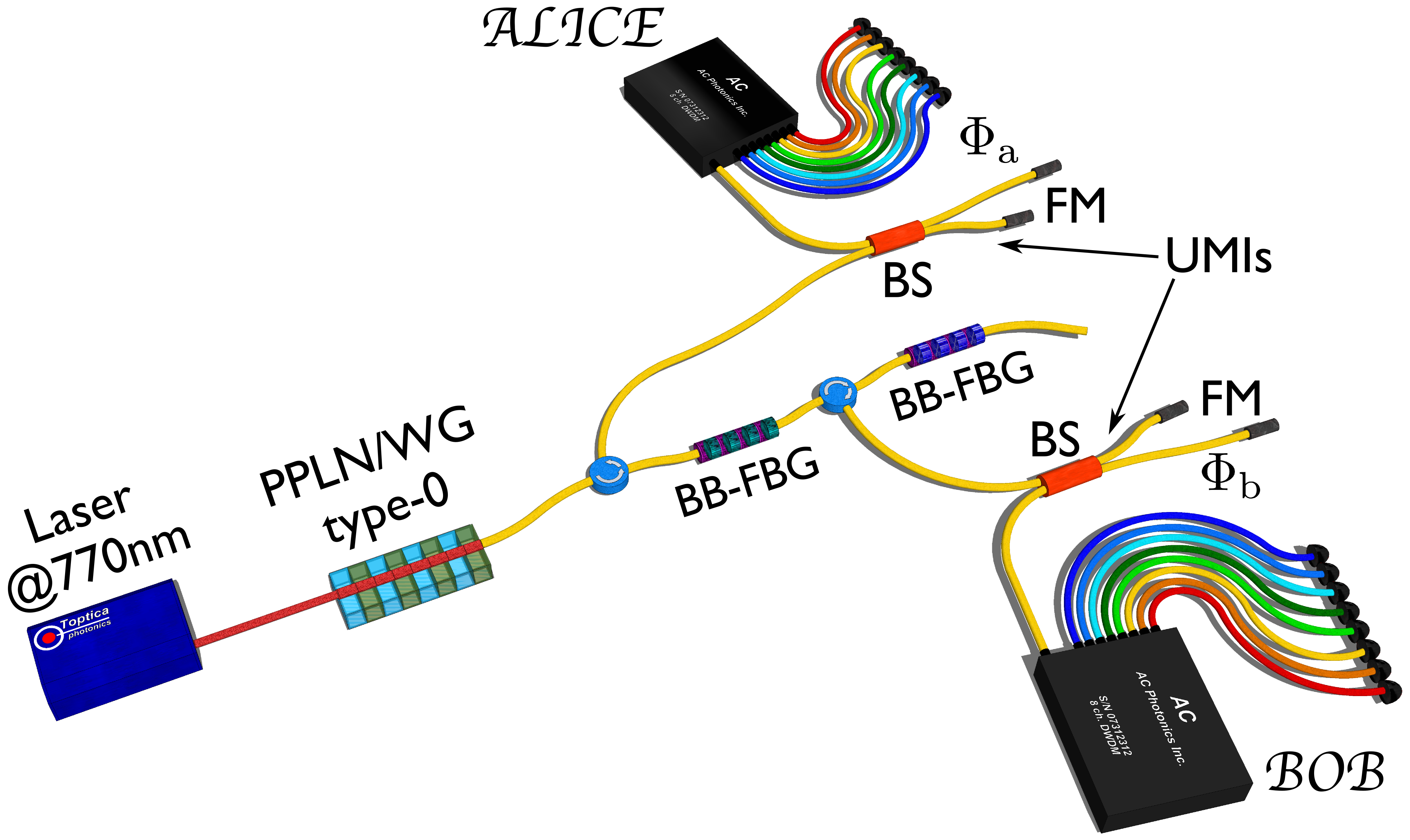}
\caption{\label{fig_setup} \textbf{Experimental setup}. A nonlinear waveguide pumped by a 770\,nm CW laser generates energy-time entangled photon pairs. Two broadband fiber Bragg grating filters (BB-FBG) and associated circulators split the photon pairs deterministically between Alice (short wavelengths) and Bob (long wavelengths). At both locations, the analysis system comprises an unbalanced Michelson interferometer (UMI) (made of a beam-splitter (BS) and two Faraday mirrors (FM)), an 8-channel DWDM, and single photon detectors. PPLN/WG: periodically poled lithium niobate waveguide.}
\end{figure}

The experimental setup is shown in Fig.~\ref{fig_setup}. A continuous-wave (CW) laser at 769.88\,nm (TOPTICA DL Pro) pumps a periodically poled lithium niobate waveguide (PPLN/WG) for generating broadband energy-time entangled pairs of photons via SPDC~\cite{TheKaiser14Long}. We exploit energy-time entangled photons as information carriers, due to their inherent invulnerability against polarization mode dispersion and drifts ~\cite{Ghalbouni_DWDM_OL_2013, hubel_high-fidelity_2007, liu_decoy-state_2010}, making this approach more robust for long distance distribution.
Energy-time entanglement presents two main features. First, the spontaneous character of SPDC prevents from knowing the emission times of the photon pairs. Second, this interaction is ruled by the conservation of the energy at the photon level, $\omega_{\rm p} = \omega_{\rm a} + \omega_{\rm b}$, where $\omega_{\rm p,a,b}$ stand for the frequencies of the pump and two generated photons, the latter being distributed to the users Alice and Bob, respectively.
At this point, let us briefly discuss the difference between CW and pulsed regime SPDC for wavelength-multiplexed purposes.
In the case of CW SPDC, each channel at Alice's location is associated, \textit{i.e.}, correlated, with a complementary channel at Bob's location. Due to strict conservation of the energy imposed by a narrow linewidth laser, all channel pairs are independent, \textit{i.e.}, orthogonal in the related Hilbert space. This permits obtaining optimal signal-to-noise ratios (SNR) for the coincidence counts due to the exploitation of the maximum channel capacity.
On the contrary, pulsed SPDC inherently implies a much broader pump spectral bandwidth. Depending on this bandwidth, strict conservation of the energy might no longer hold, having repercussions on photon pair spectral correlations. Consequently, channel cross-talk might arise, leading to reduced SNRs and overall channel capacity. To overcome this issue, particular engineering of the spectral characteristics of the full system (laser and photons) should be applied.
\begin{figure}
\centering
\includegraphics[width=72mm]{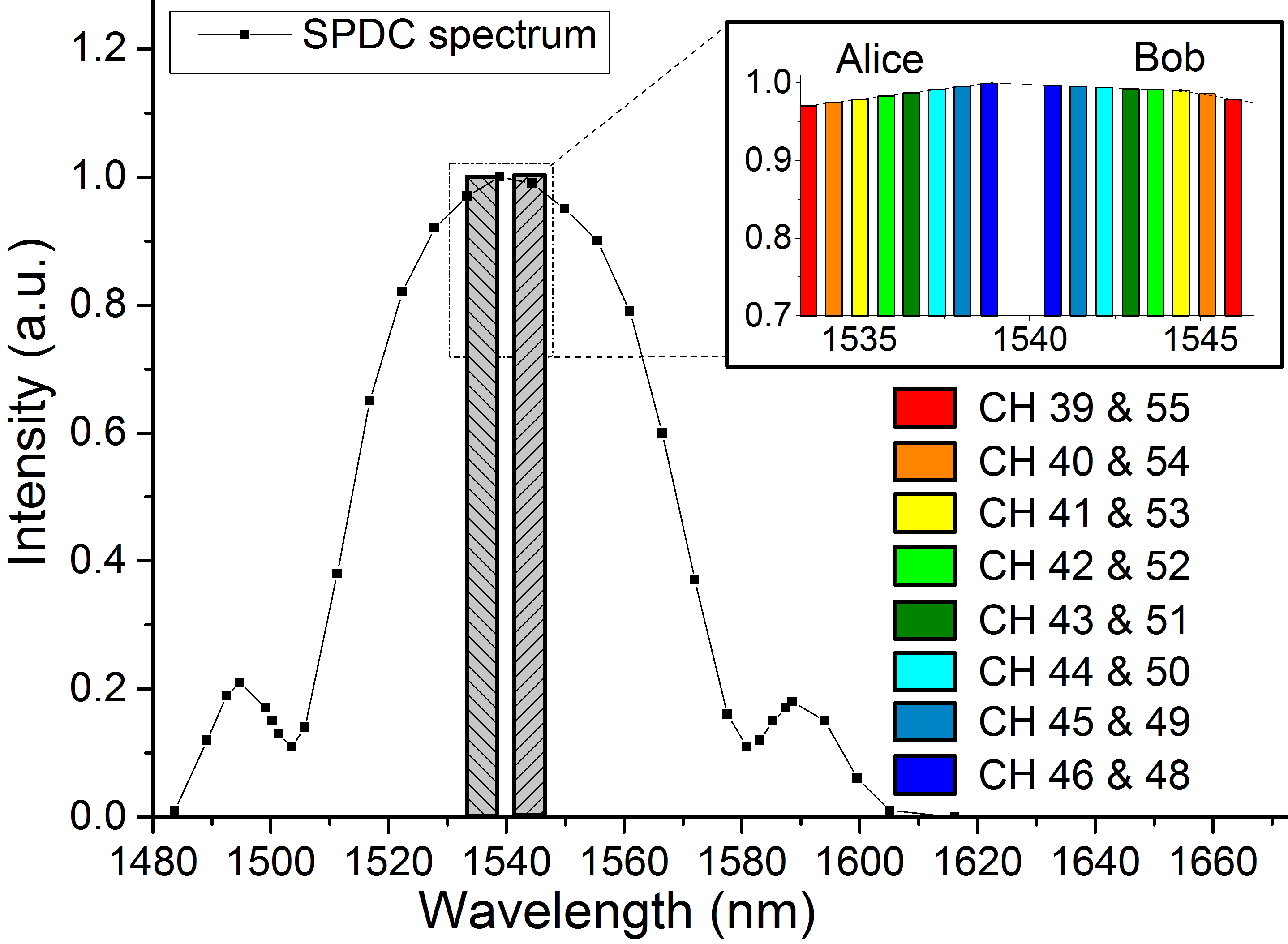}
\caption{\label{fig_spectrum} \textbf{Measured spectrum at the output of the photon pair generator}. The top-right inset shows a color coded representation of the $2 \times 8$ ITU channels used to filter the photons by pairs. On the bottom right we show the channel pairs in which entangled photon pairs are observed.}
\end{figure}
\\
For this reason, we opt for the CW regime that enables, as shown in Fig.~\ref{fig_spectrum}, generating paired photons emitted symmetrically around the degenerate wavelength of  1539.77 nm. The phase matching condition, at a crystal temperature of $\sim 120^{\circ}\rm C$, is engineered such that a $\sim 50\rm\,nm$ broadband emission spectrum is obtained for the photon pairs, covering the entire telecommunication C-band (1530-1565\,nm) and with an SPDC process efficiency of $4\cdot 10^{-6}$ photon pairs per pump photon. Such a bandwidth would allow entanglement distribution in up to 31 standard channel pairs using off-the-shelves, and high-performance multi-channel DWDM components~\cite{Aboussouan10_Qrelay}.
The emitted spectrum is directly collected thanks to a bare fiber with 55\,\% efficiency and then, \textit{ab initio}, deterministically demultiplexed so as to provide Alice and Bob with short and long wavelength photons apart from degeneracy, respectively, by means of standard broadband fiber Bragg gratings (AOS GmbH) and associated circulators. This strategy allows avoiding the 50\% loss that would arise when separating the photon pairs using a beam-splitter. Then, to reveal energy-time entanglement, we employ a set of unbalanced Michelson interferometers (UMI) in the ``Franson configuration''~\cite{Franson89}. They are made of a fiber optics beam-splitter connected to two Faraday mirrors allowing to automatically compensate polarization rotations such that excellent long term stability is guaranteed. To further exploit the potential of the broadband photon pair generator, Alice and Bob analyze entanglement, for a proof-of-concept demonstration, in 2$\times$8 complementary channels simultaneously, by demultiplexing them with standard DWDMs (AC Photonics). As shown in Fig.~\ref{fig_spectrum}, Alice is supplied with channels 39 to 46 and Bob with 48 to 55, according to the ITU grid. The total optical loss from the photon pair generator to after the DWDMs is about $5 - 6\rm \,dB$. In the end, the photons are detected using free-running indium-gallium-arsenide (InGaAs) single photon detectors. The detector at Alice's location features 440\,Hz of dark counts at 28\% detection efficiency (ID Quantique id230), while the detector at Bob's place shows 1400\,Hz at 20\% (ID Quantique id220). Both detectors are set to a dead-time of $9\,\rm \mu s$ in order to keep the probability of afterpulses low. Coincidence measurements between correlated pairs of detectors are performed using a time to amplitude converter (ORTEC 567) and related electronics. The timing jitter of the full detection system was measured to be 155 ps.

\subsection{Analysing energy-time entanglement}

We detail here the generation and analysis of energy-time entangled states. The non-normalized state at the output of the crystal reads
\begin{equation} 
\begin{split}
 |\Psi\rangle  \propto \int \int dt_{\rm a}dt_{\rm b}\psi(t_{\rm a},t_{\rm b})\\
 \times e^{-i(\omega_{\rm p}/2)(t_{\rm a}+t_{\rm b})}\widehat{a}^{\dagger}(t_{\rm a}) \widehat{b}^{\dagger}(t_{\rm b}) |0\rangle, 
 \end{split}
 \end{equation}
in which $\omega_{\rm p}$ is the pump laser frequency, $\widehat{a}^{\dagger}$ and $\widehat{b}^{\dagger}$ are the creation operators for Alice's and Bob's photons at time $t_{\rm a}$ and $t_{\rm b}$, respectively. $\psi(t_{\rm a},t_{\rm b})$ comprises both the coherence of the pump laser and the temporal shape of the biphoton wavefunction after filtering stages~\cite{AliKhanET06}. In our case, a good approximation reads $\psi(t_{\rm a},t_{\rm b})  \approx e^{-(t_{\rm a}-t_{\rm b} )^{2}/4\tau^{2}} e^{-t_{\rm a}^{2}/4T^{2}}$. Here, $\tau \approx 10\,\rm ps$ is the single photon coherence time after filtering into DWDM channels, and $T \approx 1\,\rm \mu s$ is the biphoton coherence time that corresponds to that of the pump laser~\cite{AliKhan_LargeAlpha, Kwiat1993}.

At the output of the UMIs, the two contributions for which Alice's and Bob's photon take both the short or the long path are indistinguishable. As mentioned above, this is ensured thanks to the spontaneous character of SPDC that makes it impossible to predict the generation time of the photon pairs. In other words, the detection time at Alice or Bob does not reveal the path taken by the photons.

In order to measure high quality entanglement, several experimental constraints have to be fulfilled to guarantee that these two contributions remain indistinguishable. 
All of those requirements are related to the path travel time differences in Alice's and Bob's UMIs ($\Delta \tau_{\rm a,b}$).
First of all, they must be identical within the single photon coherence time, \textit{i.e.}, $ |\Delta \tau_{\rm a}-\Delta \tau_{\rm b}  | \ll \tau$. Then, they should be greater than the coherence time $\tau$ of the filtered single photons in order to avoid single photon interference. Finally, they should be smaller than the biphoton coherence time to be able observe two photon interference. These conditions can be summarized as follow
\begin{equation}\label{eq:eq1}
T>\Delta \tau_{a,b}>\tau.
\end{equation}
In our case, we use UMIs with $\Delta \tau_{\rm a,b} = 340 \pm 10 \rm \,ps $ which fulfils all the above mentioned requirements.
As usually done, the quality of entanglement is then inferred by a two-photon correlation measurement. We choose to scan the optical phase of one of the two UMIs through temperature control which allows observing sinusoidal interference fringes of the two photon coincidence rate between Alice and Bob. To quantify the entanglement quality, a genuine figure of merit is the visibility of the obtained fringes, defined as the ratio of amplitude to average signal~\cite{Cabello09}.

\section{Results}

\subsection{Setup performance and linear scaling of the obtained coincidence rates}

We infer the quality of our entanglement-based quantum link, designed to be complient with the Ekert protocol~\cite{Ekert91} in the Franson configuration, by registering the coincidence rates between each of the 8 paired channels. The associated data are two-photon interference patterns that oscillate as a function of the sum of the phases set in Alice and Bob interformeters. Then, we deduce the entanglement quality by measuring the visibility of the obtained fringes which stands as a genuine figure of merit for both attesting entanglement and implementing secure QC.\\
Unless specified, the coincidence rates between different channel pairs are measured sequentially, \textit{i.e.}, we measure the rate in one given channel pair (typical measurement time is a few seconds), and then proceed to the next. Let us stress that the phase settings are changed only after measurements have been performed in all the eight channel pairs. In other words, the same experimental results would be obtained by employing eight pairs of detectors.\\ 
\begin{figure}
\centering
\includegraphics[width=84mm]{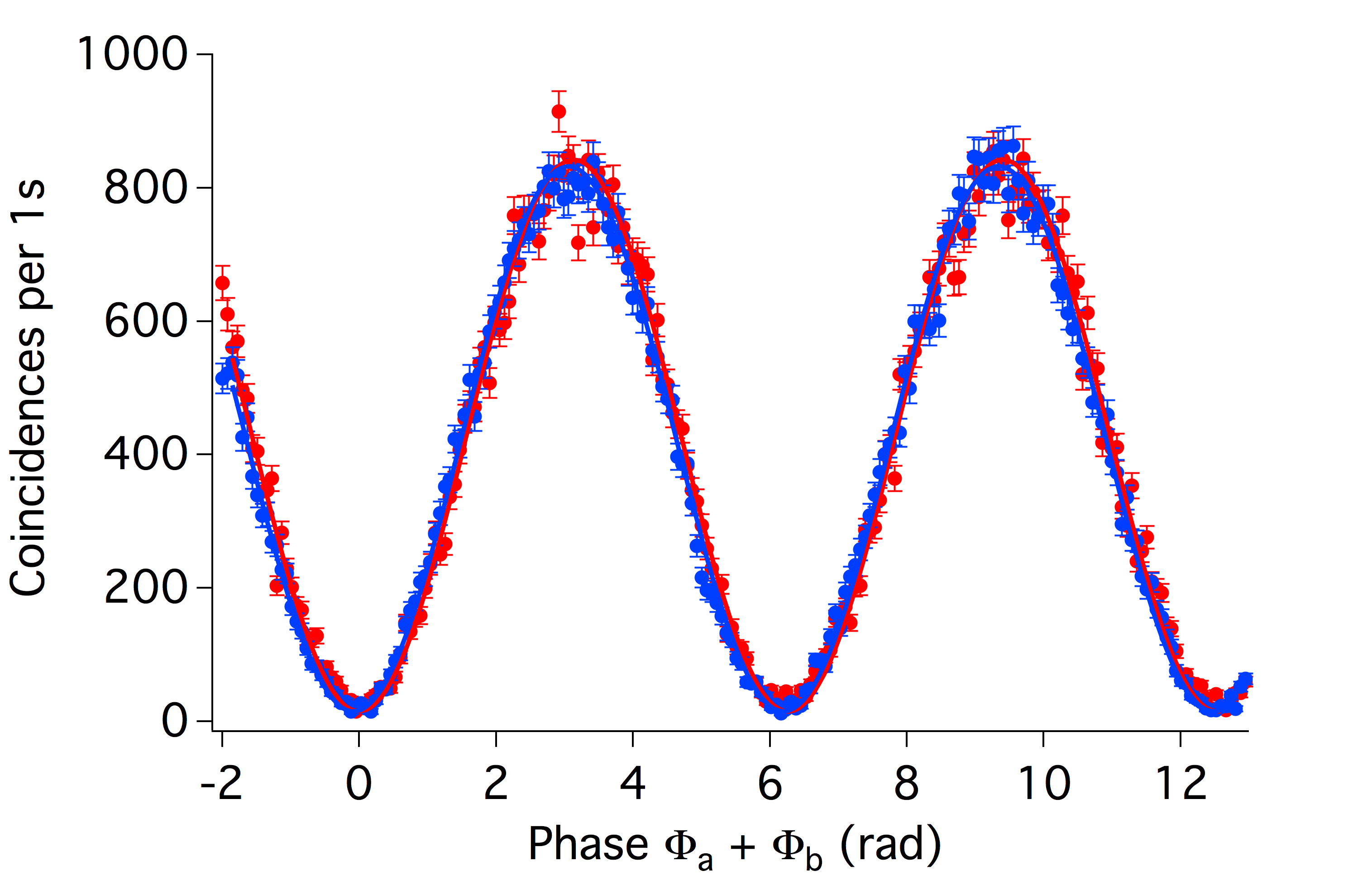}
\caption{\label{fig_interf}\textbf{Two-photon interference fringes at 0\,km distribution.} In red and blue are plotted the coincidence rates for ITU channels 39-55 (external channel pair) and 46-48 (internal channel pair), respectively. The two interference patterns are in phase thanks to nearly identical UMIs. Note that these two patterns have been obtained simultaneously, \textit{i.e.}, using two pairs of detectors.}
\end{figure}
To begin with, the pump power is set to generate a mean photon pair number $\bar{n} = 0.015$ per detection time window of 250\,ps and per pair of channels. Typical measurements for ITU channels 39-55 and 46-48 are shown in Fig.~\ref{fig_interf}.
We measure raw visibilities $V_{\rm raw}$, \textit{i.e.}, without subtraction of detector dark counts, of $96.7 \pm 1.0 \%$ and $96.5 \pm 1.1 \%$, respectively. Note that these two patterns have been recorded simultaneously using two pairs of detectors. $V_{\rm raw}$ in all the other pairs of channels is measured to be $\sim 97 \%$. It is worth mentioning that all the eight signals are in phase, which is due to the fact that both UMIs have been adjusted to identical path length differences within a few micrometers. This is almost three orders of magnitude shorter than the typical coherence length of the single photons. Additional measurements at low pump power $(\bar{n} = 0.003)$ yield raw visibilities above $98\%$, thus demonstrating the high quality of both the entanglement generator and associated measurement setup.
Note that all results are well above the threshold of $\sim71\%$ required to violate adapted Bell-Clauser-Horne-Shimony-Holt inequalities~\cite{Franson89}. Concerning practical QC links, depending on the exact protocol~\cite{Gisin02_QC,Scarani09_securityQKD}, quantum bit error rates (QBER) below 9\% are required, \textit{i.e.}, $V_{\rm raw} > 82\%$, which is also well guaranteed by our implementation.

\begin{figure}
\centering
\includegraphics[width=85mm]{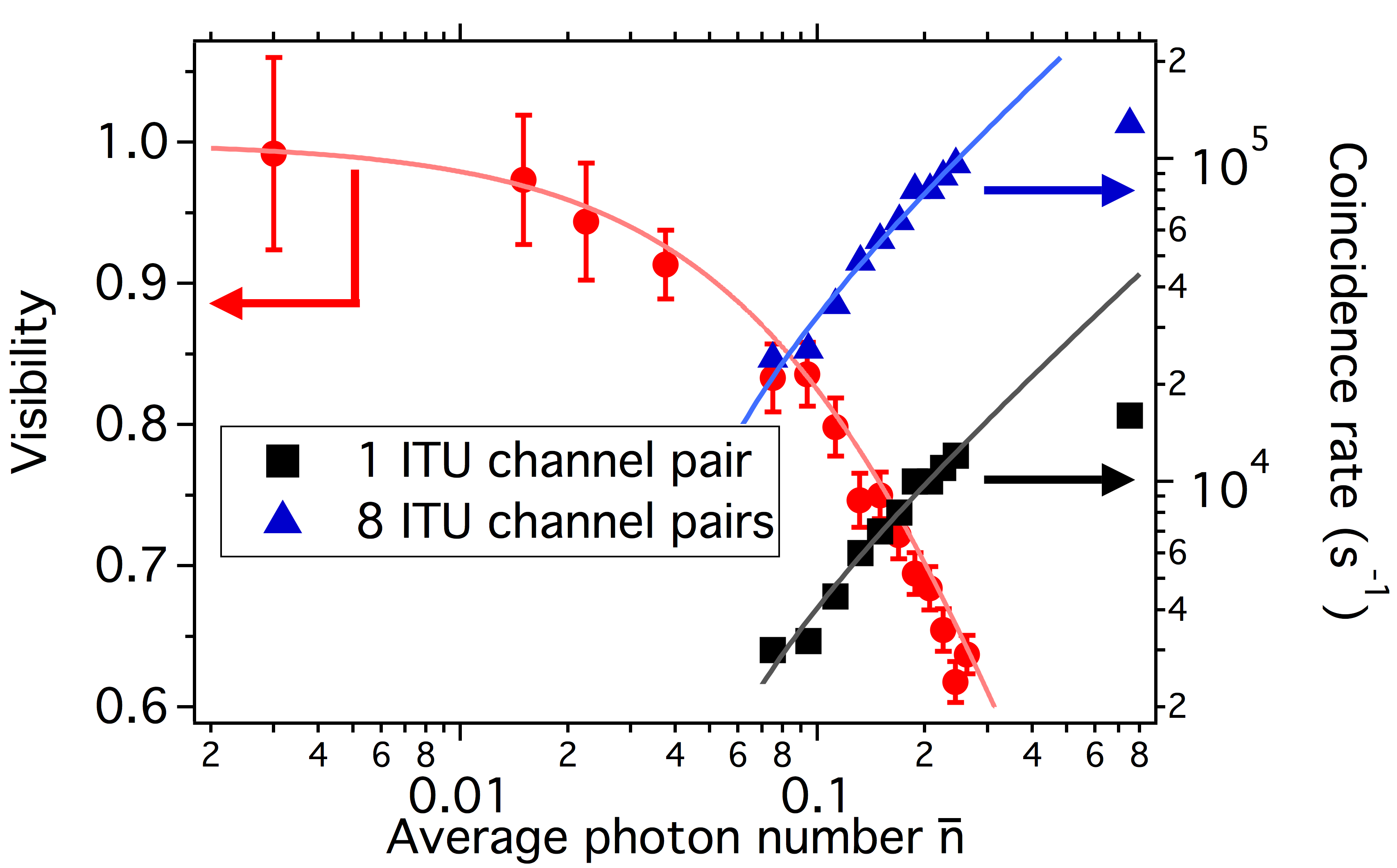}
\caption{\label{fig_vis} \textbf{Visibility and coincidence rates versus $\bar{n}$.} The experimental data (red dots) are in excellent agreement with the theoretical curve (red line) which has no free parameters, only experimentally measured ones: the losses are $\alpha_{\rm a,b} = -11 \rm \,dB$ and the dark count probabilities are $d_{\rm a,b}=2.5 \cdot 10^{-7}$ per detection window of 250\,ps.The coincidence rates scale linearly with $\bar{n}$ until detector saturation is reached at high values. For distribution in 8 ITU channel pairs, we measure an 8-fold increased total coincidence rate.}
\end{figure}

Besides exploiting the multiplexing strategy towards achieving higher bit rates~\cite{Meany_LPR14}, one might be tempted to increase $\bar{n}$ to obtain higher rates of entangled photon pairs. However, this comes at the cost of increased multiple pair event contributions, which ultimately reduce the observed entanglement visibility.
On the contrary, at too low pump powers, the visibility is also strongly reduced, as the coincidence rate drops below the noise in the detection system.
 The derived expression for the visibility, given in Ref.~\cite{Shields09_200km}, reads
\begin{equation}\label{eq:eq2}
V= \dfrac{\dfrac{\bar{n}\alpha_{\rm a}\alpha_{\rm b}}{4}}{\dfrac{\bar{n}\alpha_{\rm a}\alpha_{\rm b}}{4}+2(\bar{n}\alpha_{\rm a}/2+d_{\rm a})(\bar{n}\alpha_{\rm b}/2+d_{\rm b})}.
\end{equation}
Here, $\alpha_{\rm a,b}$ are the losses from the photon pair generator to Alice's and Bob's detectors, respectively, and $d_{\rm a,b}$ are the probabilities of having a dark count per coincidence measurement time window, in our case $d_{\rm a,b} \approx 2.5 \cdot 10^{-7}$.
As depicted in Fig.~\ref{fig_vis}, we experimentally study the evolution of the raw visibility for a wide range of pump powers, corresponding to $\bar{n}$ ranging from $0.003$ to $1$.
We observe an excellent agreement between theory and experiment for low $\bar{n}$, which constitutes our target working area. The discrepancy at higher $\bar{n}$ is explained by the fact that the theoretical model takes into account only double- and triple-pair contributions. This behaviour is also confirmed by the conclusions outlined by Lim and co-authors in Ref.~\cite{LimDiscrep}.

At this stage, we can determine the maximum tolerable $\bar{n}$ in order fulfill the above-mentioned visibility criteria.
In this perspective, operating our system at $\bar{n} = 0.10$ leads to $V_{\rm raw} \approx$ 83\% and to a two-fold coincidence rate of $\sim4.5 \cdot 10^{3}$\,s$^{-1}$ in each DWDM channel pair.
Moreover, the total coincidence rate between Alice and Bob, \textit{i.e.}, taking into account the 8 correlated pairs of channels, reaches $\sim 3.8 \cdot 10^{4}$\,s$^{-1}$ and clearly demonstrates, as shown in Fig.~\ref{fig_vis} the linear scaling as a function of the number of exploited pairs of correlated channels. Note that our setup is stable enough to collect coincidences over hours of measurement at rates up to $1.4 \cdot 10^8$\,events$/$hour.

\subsection{Entanglement distribution over 150\,km}

We now study the performance of our system in a long distance scenario. To this end, we distribute the photons over a 150\,km standard fiber link made of two SMF28e fiber spools and adapted dispersion compensating fiber (DCF) modules, thereby mimicking a realistic implementation of an actual fiber quantum network. According to equation~\ref{eq:eq2}, we can tolerate up to 47\,dB of loss in each channel while keeping $V_{\rm raw}>82\%$. This corresponds to a fiber communication link spanning over $\sim \rm 360\,km$. However, in this case, we would only register a few coincidences per hour.
Consequently, we opt for entangled photon pair distribution over 150\,km, which stands as a trade-off between practical coincidence rates and significant optical fiber length giving losses measured to be about 32\,dB.

\begin{figure}
\centering
\includegraphics[width=84mm]{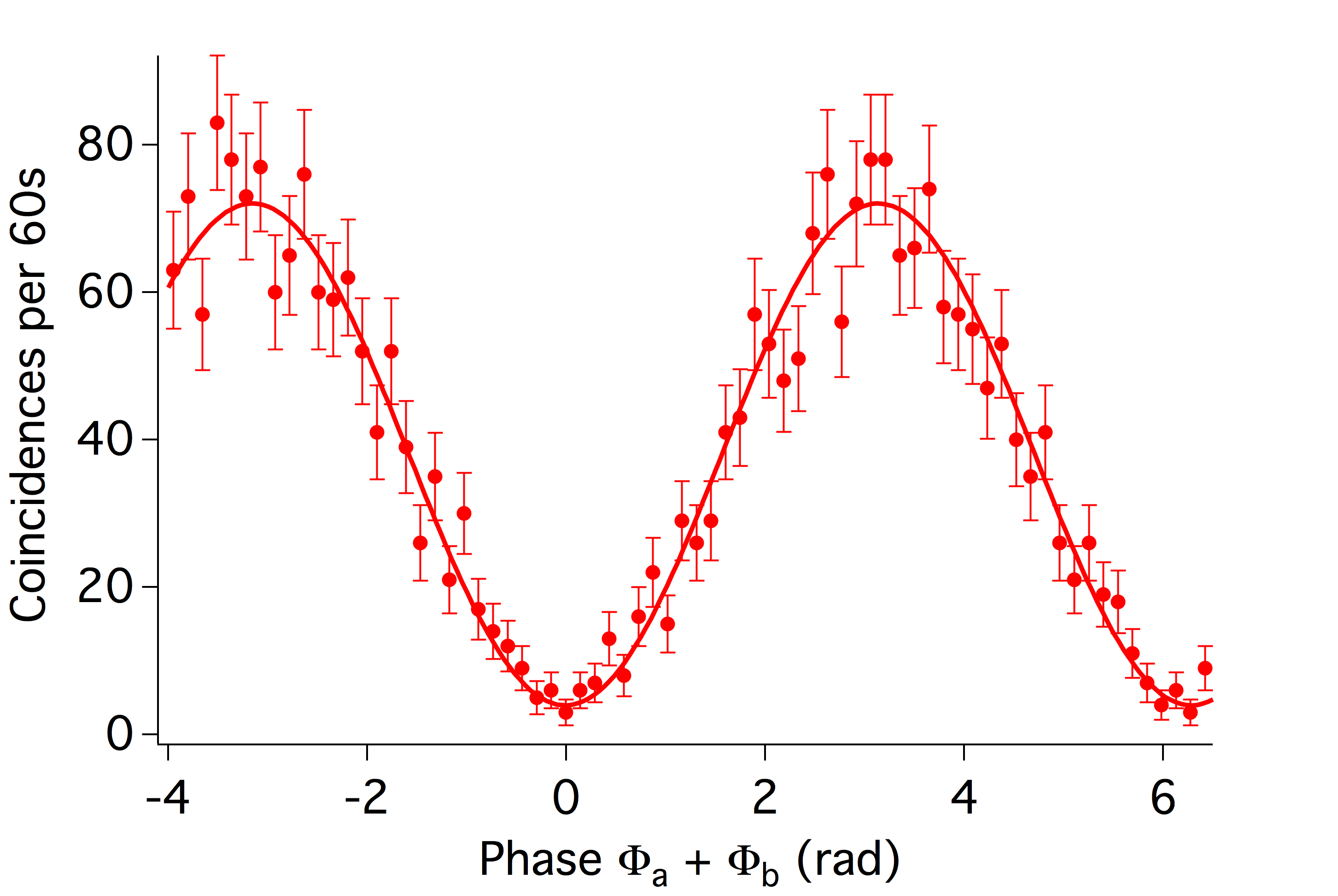}
\caption{\label{fig_interf150}\textbf{Two-photon interference fringes at 150\,km distribution.} Coincidence rates are shown for the channel pair 45-49.}
\end{figure}

Thanks to our flexible setup, we can operate the source at various $\bar{n}$ so as to choose, in real-time and as a function of the experienced signal-to-noise ratio (which notably depends on the distance between the users), the suitable compromise between bit rate and security in a QC scenario. For this particular configuration, we choose $\bar{n}=0.05$, for which we expect $V_{\rm raw}\sim$90\%. Fig.~\ref{fig_interf150} (red curve) shows exemplary visibility measurements in the paired channels 45-49. We obtain a two-fold coincidence rate of $\sim$1.1\,s$^{-1}$ and a raw visibility of $89.4 \pm 3.5\%$. Correspondingly, the average visibility obtained in all 8 channel pairs is measured to be $\sim 87\%$, for a total coincidence rate of $\sim$9\,s$^{-1}$.

This result means that almost one order of magnitude higher coincidence rates can be obtained compared to previously reported experiments having similar configurations~\cite{Shields09_200km,Takesue13_300km,Takesue10,Lim10}.

\section{Discussion}

We present in Table~\ref{tab:table1} a performance comparison of our work with pertinent realizations that have been reported in the literature~\cite{Takesue13_300km,Lim10,Herbauts_DWDM_OPEX_2013,Ghalbouni_DWDM_OL_2013}. 
\hspace{-3cm}
\renewcommand{\arraystretch}{0.6}
\begin{table*}
\begin{center}
\begin{threeparttable}
\vspace{2mm}
\begin{tabular}{c c c c c c c c c}
\hline
\hline
{}&\multicolumn {3}{c} {Quantum fiber link}&\multirow{2}*{Observable}&\multirow{2}*{Protocol}&\multirow{2}*{Detector}&\multirow{2}*{$C \rm \, (s^{-1})$}&\multirow{2}*{\textit{V}(\%)}\\
\cline{2-4} {}&\textit{L}(km)&Fiber&Regime \\
\hline
This work & 150 & Standard & CW & ET & Eckert91 & InGaAs & 9 & 87\\
Inagaki (2013)~\cite{Takesue13_300km} & 300 & DSF & Pulsed & TB & Eckert91 & SNSPD & 0.02&84\\
Takesue (2010)~\cite{Takesue10} & 100 & DSF & Pulsed & TB & Eckert91 & InGaAs & 4.8 &\textit{N/A}\\
Lim (2010)~\cite{Lim10} & 10 & Standard & CW & Polar & Eckert91 & InGaAs & 2 & 87\\
H\"{u}bel (2013)~\cite{hubel_high-fidelity_2007} & 0 & Standard & CW & Polar & Eckert91 & InGaAs & 450 & 89\\
Korzh (2015)~\cite{Korzh_QKD300km_2015} & 307 & ULL & Pulsed & Phase & COW & InGaAs & 3.18 & \textit{N/A}\\
Shimizu (2014)~\cite{shimizu_performance_2014} & 90 & Standard$^\dagger$ & Pulsed & Phase & DPS & SNSPD & 1100 & \textit{N/A}\\
Liu (2010)~\cite{liu_decoy-state_2010} & 200 & Standard & Pulsed & Polar & BB84 & SNSPD & 15 & \textit{N/A}\\
\hline
\hline
\end{tabular}
	\begin{tablenotes}   
    \item[$\dagger$] Installed fiber link  
	\end{tablenotes} 
\end{threeparttable}
\caption{Comparison of notable QKD demonstrations used for long distance distribution. Meaning of the acronyms -- \textit{L}: length; \textit{C}: total coincidence rate; \textit{V}: visibility; TB, ET, and Polar: time-bin, energy-time, and polarization observable, respectively; CW: continuous wave; DSF: dispersion-shifted fiber; ULL: ultra low loss; COW: coherent-one-way protocol; DPS: differential phase shift protocol; BB84: Bennett \& Brassard protocol with decoy-state pulses; SNSPD: superconducting nanowire single photon detector; InGaAS: indium-gallium-arsenide single photon detector; \textit{N/A} : (data) not available.\label{tab:table1}}
\end{center}
\end{table*}

It is worth noticing that the level of performance demonstrated in previous realizations ~\cite{Takesue13_300km,Takesue10,shimizu_performance_2014,liu_decoy-state_2010,Korzh_QKD300km_2015} (see Table~\ref{tab:table1}) relies either on technological advances such as efficient and low dark-counts detectors, or non-standard fibers, or on high repetition rates. On the contrary, our demonstration only takes advantage of standard  telecom components towards reaching a higher level of performance. 
Compared to the experiment reported by Lim and co-workers~\cite{Lim10}, in which polarization entanglement is analysed sequentially at different wavelengths, we believe that energy-time entanglement is much better suited for this kind of applications as the associated analysis system is made of stand-alone devices. In other words, the latter do not necessitate any phase stabilization as a function of the wavelength~\cite{Xavier_11}, making it possible to analyze entanglement simultaneously in all the correlated channel pairs.
 Furthermore, other protocols could also benefit from such an approach, \textit{e.g.}, those exploring high-dimensional QC towards increasing the number of encoded bits per photon~\cite{HD_QKDDispersive13,AliKhan_LargeAlpha}.

\section{Conclusion \& outlook}

As a conclusion, the use of a broadband entanglement generator combined with off-the-shelves 8-channel DWDM components allowed us to show that coincidence rates scale linearly with the number of exploited correlated pairs of channels. By introducing entanglement distribution into a long-distance and wavelength multiplexed environment, we have successfully realised an essential building block towards the future realization of a quantum network, with the potential of being compliant with device-independent strategies~\cite{Vazirani_PRL_14,Hensen_LoopholeFree_15}.  We also note that for an actual quantum key distribution implementation, both users would need to employ an additional interferometer for quantum state analysis in the complementary basis.
Our realization is the first in which a single photon pair generator enables the simultaneous distribution of photonic entanglement over 8 independent DWDM pairs of channels. 
We also stress the potential of this approach as there exist various strategies to further increase the performance of our system. As for classical communication, the maximal channel capacity would be reached if the single photon spectral bandwidth was similar to the detection bandwidth. Our system's capacity can therefore straightforwardly be scaled up by using a denser channel spacing and/or by increasing the detection bandwidth (which is the inverse of the detector's timing resolution).
For example, current commercially available 12.5\,GHz or 25\,GHz multichannel ultra-DWDMs would be able to match the detection bandwidth of superconducting single photon detectors, typically on the order of a few tens of GHz~\cite{Hadfield09}.
Another consideration is the number of employed DWDM channels. In this work, we use $2 \times 8$ channels, however, the telecommunication C-Band covers $2 \times 22$ channels in the 100\,GHz grid ($2 \times 176$ channels in the 12.5\,GHz grid). In such a broadband scenario, one has to circumvent fiber chromatic dispersion, which usually limits the number of in-phase pairs of channels. One way to do so would be to properly unbalance the UMI analyzers such that the entanglement can be exploited simultaneously over the entire telecom C-band. This is, in our case, straightforwardly achievable and will be reported elsewhere. Moreover, with such particular devices, coincidence rate could be augmented up to 200\,bits/s after 150\,km distribution.
\\

\noindent
{\textbf{Acknowledgements}} The authors thank C. Gonnet and P. Sansonetti from the Prysmian Group for lending long standard fibers and dispersion compensation fiber modules. Financial support from the ``Agence Nationale de la Recherche'' for the \textsc{Conneqt} (ANR-2011-EMMA-0002) and \textsc{Spocq} (ANR-14-CE32-0019) projects, the European FP7-ITN PICQUE project (grant agreement N$^{o}$ 608062), the iXCore Research Foundation, and the Simone \& Cino Del Duca Research Foundation (Institut de France), is acknowledged.



\end{document}